\documentclass[aps,prd,reprint,twocolumn,superscriptaddress,showpacs]{revtex4-1}

\usepackage{graphicx}
\usepackage{mathrsfs}
\usepackage{bm}
\usepackage{amsmath}
\usepackage{dcolumn}
\usepackage{dsfont}
\usepackage{amssymb}
\usepackage{tabularx}
\usepackage{array}
\usepackage{float}
\usepackage{color}
\usepackage{epstopdf}
\usepackage{mathrsfs}
\usepackage[colorlinks, linkcolor=blue,anchorcolor=blue,citecolor=blue,urlcolor=blue]{hyperref}

\usepackage[mathlines]{lineno}
\usepackage{bbold}
\usepackage[T1]{fontenc}
\usepackage{multirow}

\newcommand{\angstrom}{\text{\normalfont\AA}}

\graphicspath{{./}{figure/}}

\begin{document}

\title{Highly anisotropic two-dimensional metal in monolayer MoOCl$_2$}

\author{Jianzhou Zhao}
\affiliation{Co-Innovation Center for New Energetic Materials, Southwest University of Science and Technology, Mianyang 621010,
China}
\affiliation{Research Laboratory for Quantum Materials, Singapore University of Technology and Design, Singapore 487372, Singapore}

\author{Weikang Wu}
\affiliation{Research Laboratory for Quantum Materials, Singapore University of Technology and Design, Singapore 487372, Singapore}

\author{Jiaojiao Zhu}
\affiliation{Research Laboratory for Quantum Materials, Singapore University of Technology and Design, Singapore 487372, Singapore}

\author{Yunhao Lu}
\affiliation{Zhejiang Province Key Laboratory of Quantum Technology and Device, Department of Physics, Zhejiang University, Hangzhou 310027, China}

\author{Bin Xiang}
\affiliation{Hefei National Research Center for Physical Sciences at the Microscale, Department of Materials Science \& Engineering,
University of Science and Technology of China, Hefei 230026, China}

\author{Shengyuan A. Yang}
\affiliation{Research Laboratory for Quantum Materials, Singapore University of Technology and Design, Singapore 487372, Singapore}

\begin{abstract}
Anisotropy is a general feature in materials. Strong anisotropy could lead to interesting physical properties and useful applications. Here, based on first-principles calculations and theoretical analysis, we predict a stable two-dimensional (2D) material---the monolayer MoOCl$_2$, and show that it possesses intriguing properties related to its high anisotropy. Monolayer MoOCl$_2$ can be readily exfoliated from the van der Waals layered bulk, which has already been synthesized. We show that a high in-plane anisotropy manifests in the structural, phononic, mechanical, electronic, and optical properties of monolayer MoOCl$_2$. The material is a metal with highly anisotropic Fermi surfaces, giving rise to open orbits at the Fermi level, which can be probed in magneto-transport. Remarkably, the combination of high anisotropy and metallic character makes monolayer MoOCl$_2$ an almost ideal hyperbolic material. It has two very wide hyperbolic frequency windows from 0.41 eV (99 THz) to 2.90 eV (701 THz), and from 3.63 eV (878 THz) to 5.54 eV (1340 THz). The former window has a large overlap with the visible spectrum, and the dissipation for most part of this window is very small. The window can be further tuned by the applied strain, such that at a chosen frequency, a transition between elliptic and hyperbolic character can be induced by strain. Our work discovers a highly anisotropic 2D metal with extraordinary properties, which holds great potential for electronic and optical applications.
\end{abstract}

\maketitle

\section{\label{sec: intro} Introduction}

Two-dimensional (2D) materials have been attracting tremendous interest because of their fascinating properties, excellent tunability, and promising applications~\cite{doi:10.1146/annurev-matsci-070214-021034, Novoselov10451, doi:10.1021/acsnano.5b05556, doi:10.1021/acs.chemrev.6b00558, Bhimanapati2019}. Similar to 3D crystals, most 2D materials are stabilized in high-symmetry structures, which are usually preferred for minimizing the energy. For example, graphene~\cite{Novoselov666}, hexagonal BN~\cite{Nagashima1995}, H- and T-phase transition metal dichalcogenides~\cite{PhysRevLett.105.136805,eda2011}, and most MXenes~\cite{doi:10.1002/adma.201102306} have threefold rotational symmetry. It follows that their properties should be isotropic in the 2D plane.

There do exist 2D anisotropic materials, such as the black phosphorene~\cite{Li.2014uga,Ling:2015ba,PhysRevB.92.165406}, the similarly structured group-V monolayers~\cite{Yunhao2016,PhysRevB.93.085424,Rehman2020}, and 2D group-IV monochalcogenides~\cite{Chang2016}. It was realized that the anisotropy could be an advantage of these materials for controlling the directional transport of charge carriers or plasmonic waves~\cite{fei2014,guan2018,Nemilentsau.2016}, for generating ferroelectricity~\cite{PhysRevLett.117.097601,xiao2018}, for tuning light absorption with selective polarization~\cite{qiao2014}, and etc.

Particularly, it was proposed that highly anisotropic metals could be good candidates for hyperbolic materials~\cite{sun2014,Guan.2017}, which means the real parts of their permittivity tensor elements take different signs~\cite{10.1038/nphoton.2013.243,PhysRevLett.90.077405,10.1155/2012/267564}. Such materials have a special hyperbolic-type light dispersion, which can produce extraordinary optical properties, like all-angle negative refraction~\cite{Smith.2004,Hoffman.2007,Yao.2008}, sub-wavelength imaging~\cite{Liu.2007,Lu.2008,Rho.2010}, and strongly enhanced spontaneous emission~\cite{Jacob.2012,Yang.2012,Cortes.2012}. The idea was explored in several 3D layered materials~\cite{sun2014,Sun.20116df,Alekseyev.2012,Esslinger.2014,Korzeb.2015}, and the class of electride compounds such as Ca$_2$N was predicted to be ideal candidates because of their wide hyperbolic frequency window and suppressed dissipation~\cite{Guan.2017}. For 2D, the aforementioned anisotropic materials such as black phosphorene could work, but they require additional charge doping~\cite{Nemilentsau.2016}. Besides, the anisotropic is not so strong, which limits the hyperbolic frequency window, and the sizable dissipation due to light absorption is also an issue. The situation therefore calls for an exploration of new 2D materials, which are intrinsically metallic and possess a high anisotropy.

MoOCl$_2$ is a typical van der Waals (vdW) layered material. In a recent experiment~\cite{Wang.2020p6}, it was found that the material is a good metal with strong anisotropy in the layer plane, and thin flakes can be easily exfoliated from the bulk sample. Motivated by this finding, in this work, we perform a systematic study of MoOCl$_2$ in the 2D limit, based on the first-principles calculations. We show that monolayer MoOCl$_2$ is stable and has low binding energy to the bulk. The lattice structure has a strong in-plane anisotropy, with binding much stronger in one direction than the other. We demonstrate that this strong anisotropy manifests in various physical properties, including the sound speed, the Young's modulus, the electronic band dispersion, the Fermi surface geometry, the optical properties, and the strain response. Remarkably, we find that monolayer MoOCl$_2$ is an almost ideal 2D hyperbolic material. It has two wide hyperbolic windows from 0.41 eV (99 THz) to 2.90 eV (701 THz) and from 3.63 eV (878 THz) to 5.54 eV (1340 THz). The former window overlaps with a large portion of the visible light. Moreover, for most part of this window, the absorption is very small. Our work reveals extraordinary properties of monolayer  MoOCl$_2$, which hold great potential for nanoscale electronic, mechanic, and optical applications.


\section{\label{sec: method} Calculation Method}

Our first-principles calculations were based on the
density functional theory (DFT), as implemented in the Vienna ab-initio Simulation Package (VASP)~\cite{Kresse1996kl,Kresse1996vk}.
The projector augmented wave (PAW) pseudo-potentials were adopted in the calculation~\cite{Kresse1999wc,Blochl1994zz}.
The generalized gradient approximation with the Perdew-Burke-Ernzerhof (PBE) realization~\cite{Perdew1996iq} was used for the exchange-correlation functional. The valence electrons treated in the calculations include Mo ($4d^55s^1$), O ($2s^22p^4$), and Cl ($3s^23p^5$). The kinetic energy cutoff was fixed to 500 eV.
For the self-consistent calculations, the Brillouin zone (BZ) integration was performed on a $\Gamma$-centered mesh of $21 \times 12 \times 1$ $k$-points. The energy and force convergence criteria were set to be 10$^{-7}$ eV and 0.001 eV$/\angstrom$, respectively. A vacuum layer with 40 $\angstrom$ thickness was taken to avoid artificial interactions between periodic images, and this layer was subtracted when obtaining the proper permittivity of the monolayer~\cite{Guan.2015,Laturia.2018}.
The optical properties were calculated by the FP-LAPW WIEN2k package~\cite{wien2k}. The plane wave cut-off parameter $K_\text{max}$ was set by $R_\text{mt}\times K_\text{max}=7.0$. The \textit{k}-point mesh for the self-consistent loop was the same as the VASP input, and it was increased to $63\times 36\times 1$ for the optical conductivity calculation.
The phonon dispersion was calculated by using a 3$\times$3 supercell. The force constants were calculated by VASP, and the pre-process and the post-process were performed by using the {Phonopy} code~\cite{phonopy, PhysRevB.81.174301}. Non-analytical term correction (NAC) was considered in treating the long range interaction of macroscopic electric field induced by the polarization of the collective ionic motions near the $\Gamma$-point~\cite{Gonze.1994,Gonze.1997}.

\section{\label{sec: structure} Crystal structure, phonon, and mechanical properties}

\begin{figure}[tb]
    \includegraphics[width=0.48\textwidth,angle=0]{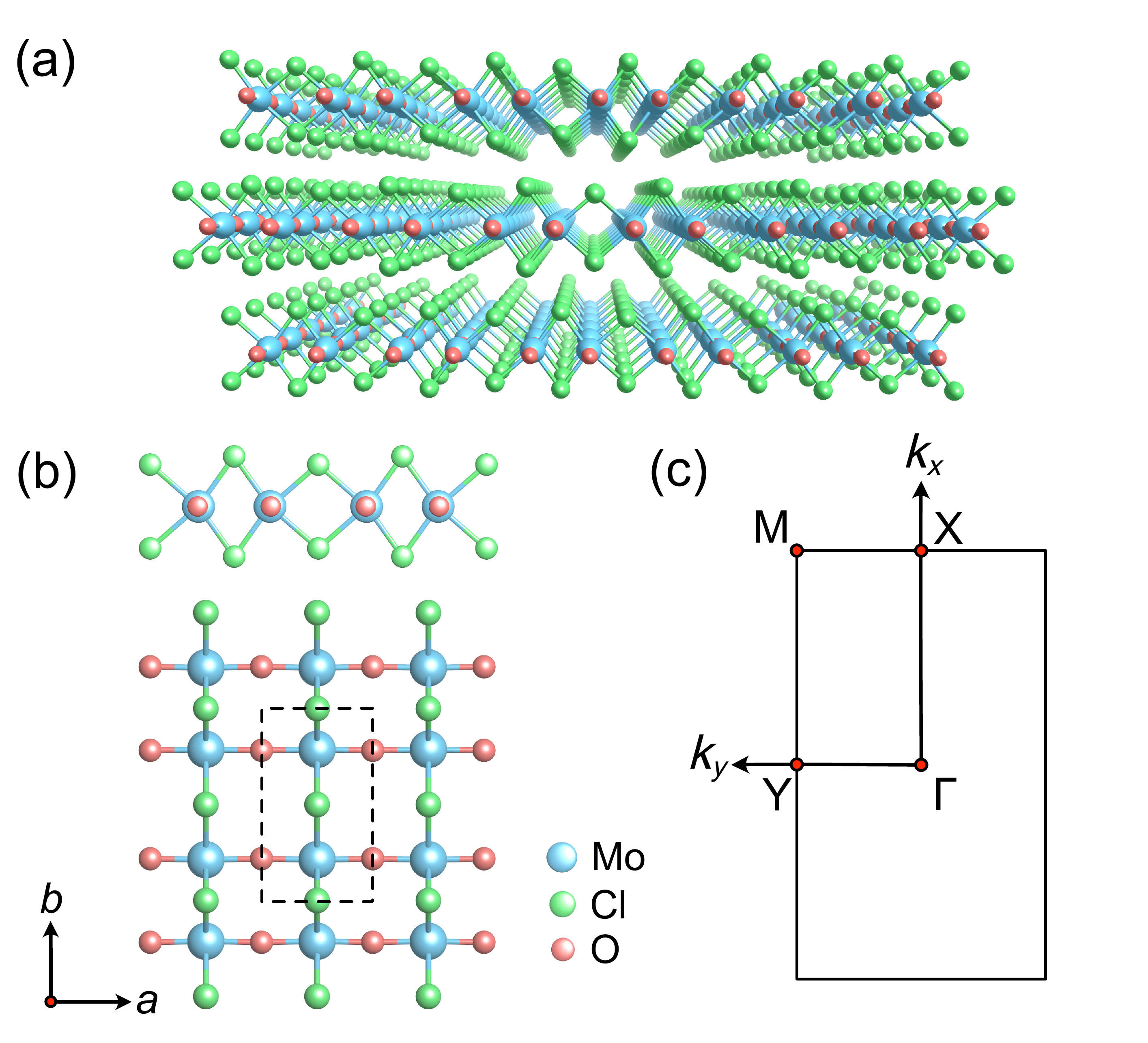}
    \caption{\label{fig: lattice} (a) Crystal structure of bulk MoOCl$_2$. (b) Side and top views of monolayer MoOCl$_2$. (c) Brillouin zone of monolayer MoOCl$_2$. The crystal structures were illustrated by using the VESTA software~\cite{Momma:db5098}.}
\end{figure}

The experimental realization of bulk MoOCl$_2$ was reported  more than 50 years ago, by using the chemical vapor transport method~\cite{SCHAFER1964152}. Bulk MoOCl$_2$
is a vdW layered material with  NbOCl$_2$-type crystal structure (space group $C2/m$), as depicted in Fig.~\ref{fig: lattice}(a). Each layer in the structure consists of three atomic layers: the central layer contains Mo and O atoms, whereas the upper and the lower layers are of Cl atoms. The properties of bulk MoOCl$_2$ have been studied in Ref.~\cite{Wang.2020p6}. In this work, we shall focus on MoOCl$_2$ in the monolayer form.

Because of the weak vdW bonding, we expect that a single layer of MoOCl$_2$ can be readily exfoliated from the bulk sample. To have a quantitative assessment, we calculate the interlayer binding energy, defined as
\begin{equation}
  E_b = (E_\text{ML} + E_{n-1} - E_{n})/A,
\end{equation}
where $E_\text{ML}$, $E_{n-1}$, and $E_n$ are the energies of monolayer, $(n-1)$ layers, and $n$ layers, respectively, and $A$ is the in-plane area. Formally, $E_b$ should take the limiting value when $n$ approaches infinity. In practice, for vdW layered materials, $E_b$ typically has a quick saturation after first few layers. In our calculation, we take $n=9$, where the convergence is achieved within 0.1 meV/\AA$^2$. The obtained binding energy for monolayer MoOCl$_2$ is 18.0 meV/\AA$^2$. This value is comparable to graphene~\cite{PhysRevB.85.205418,10.1038/ncomms8853} ($\sim 12$ to $24$ meV/\AA$^2$) and monolayer MoS$_2$~\cite{PhysRevMaterials.3.044002} ($\sim 15$ meV/\AA$^2$), suggesting that monolayer MoOCl$_2$ can also be obtained via mechanical exfoliation from the bulk sample.

The fully relaxed structure of monolayer MoOCl$_2$ is illustrated in Fig.~\ref{fig: lattice}(b). The structure can be viewed as consisting of parallel Mo-O chains running along the $x$ ($a$ vector) direction, and these chains are connected by the Cl atoms. It has a rectangular lattice with space group $Pmmm$. The symmetry elements in the group include the twofold rotations $C_{2i}$  along the three orthogonal direction, the inversion, and three mirror planes $M_i$ ($i=x,y,z$). Clearly, the structure is anisotropic in the 2D plane. The optimized lattice constants are given by $a=3.805$ \AA\ and $b=6.561$ \AA. The detailed atomic positions are listed in Table~\ref{tab: lattice}.

\begin{table}[h]
    \caption{\label{tab: lattice} Atomic positions in monolayer MoOCl$_2$ obtained from our calculation. Here, the values are given in the form of fractional coordinates. The lattice parameter along $z$ is taken to be 40 \AA.}
    \begin{ruledtabular}
        \begin{tabular}{ccccc}
            Site         & Wyckoff symbol & $x$        & $y$       & $z$        \\
            \hline
            Mo$_1$       & 2$p$             & 0.500    & 0.787   & 0.500    \\
            O$_1$        & 2$n$             & 0.000    & 0.218   & 0.500    \\
            Cl$_1$       & 2$s$             & 0.500    & 0.000   & 0.467    \\
            Cl$_2$       & 2$t$             & 0.500    & 0.500   & 0.473    \\
        \end{tabular}
    \end{ruledtabular}
\end{table}

\begin{figure}[h]
    \includegraphics[width=0.48\textwidth,angle=0]{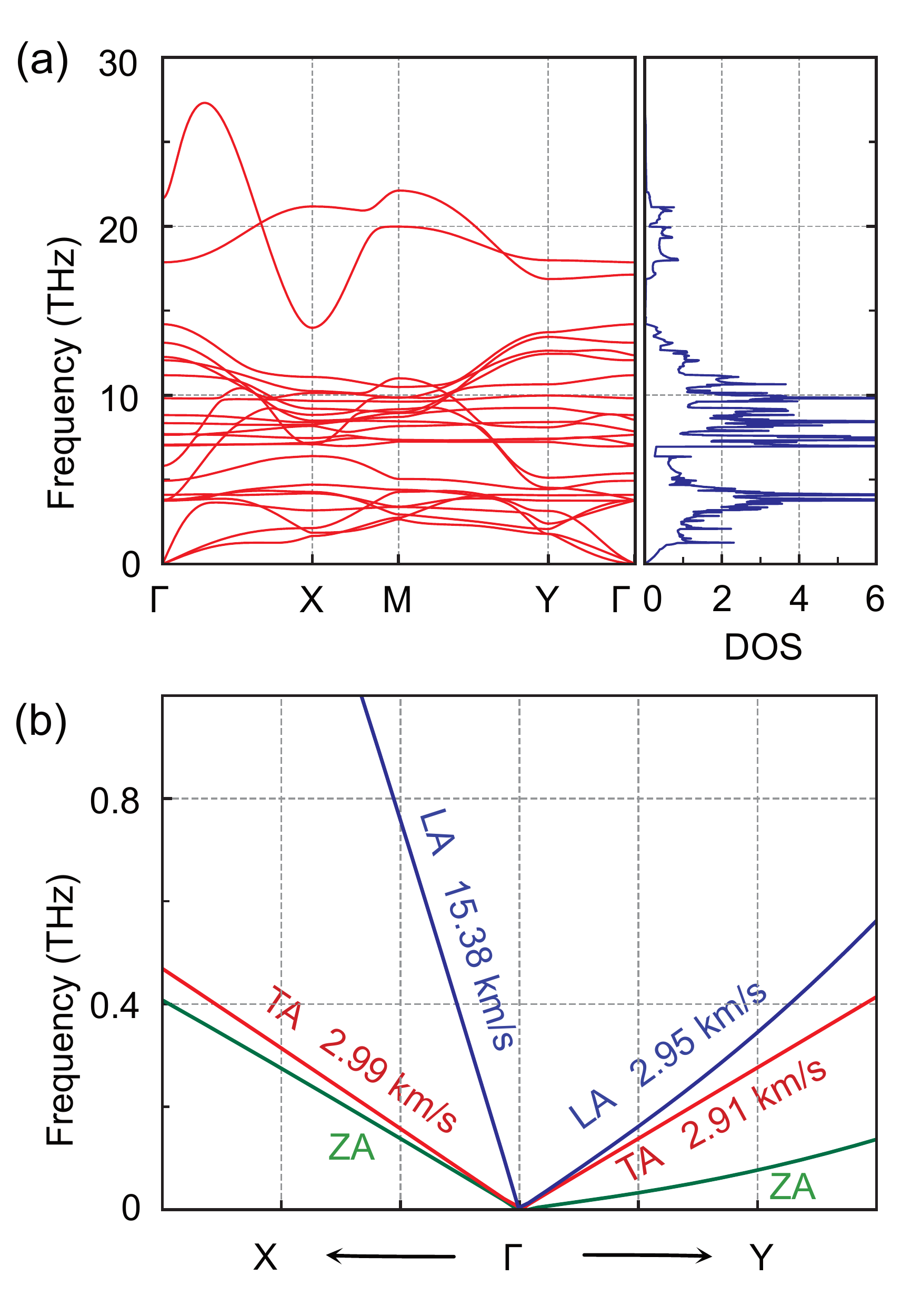}
    \caption{\label{fig: phonon} (a) Phonon spectrum and phonon DOS for monolayer MoOCl$_2$. (b) Dispersion of the three acoustic branches near the $\Gamma$ point.}
\end{figure}

To investigate the stability of monolayer MoOCl$_2$, we calculated its phonon spectrum. The obtained result is plotted in Fig.~\ref{fig: phonon}(a). One observes that there is no soft mode in the spectrum, indicating that the structure is dynamically stable. Around the $\Gamma$ point and at low frequencies, one finds the three acoustic phonon branches (see Fig.~\ref{fig: phonon}(b)). The lower branch is for the out-of-plane acoustic (ZA) phonon modes, which typically has a quadratic dispersion and is regarded as a characteristic for 2D materials~\cite{Liu.20070z,Zhu.2014}. The other two branches correspond to the longitudinal acoustic (LA) and transverse acoustic (TA) modes, which have linear dispersions. Their slopes give the sound velocities in the lattice. From the result in Fig.~\ref{fig: phonon}(b), we find that the LA and TA phonon velocities along the $x$ direction (along the $a$ vector) are 15.35 km/s and 2.99 km/s, whereas the corresponding values for the $y$ direction (along the $b$ vector) are 2.95 km/s and 2.91 km/s. One observes that the LA phonon velocity along $x$ is more than 5 times larger than that along $y$. As the LA phonon velocity reflects the bonding strength in that direction, this large anisotropy indicates that the monolayer MoOCl$_2$ structure is strongly coupled along $x$, i.e., along the Mo-O chains (for reference, the LA phonon velocity in graphene is 19.9 km/s~\cite{CONG201919}), whereas the coupling along $y$, i.e., between the Mo-O chains, is much weaker.

\begin{table*}[tb]
\caption{\label{tab: ec} Elastic constants ($C_{ij}$), Young's Modulus ($Y$) and Poisson's ratio ($\nu$) of monolayer MoOCl$_2$. The standard Voigt notations are used, namely, 1-$xx$, 2-$yy$ and 6-$xy$. The results for graphene are also presented as a reference.}
\begin{ruledtabular}
\begin{tabular}{lcccccccc}
    \multirow{2}{*}{\ } & \multicolumn{4}{c}{Elastic constants (N/m)} & \multicolumn{2}{c}{Young's modulus (N/m)} & \multicolumn{2}{c}{Poisson's ratio} \\
                        & $C_{11}$ & $C_{22}$ & $C_{12}$ &  $C_{66}$   & $Y_{10}$      &   $Y_{01}$              & $\nu_{10}$ & $\nu_{01}$    \\
    \hline
    ML-MoOCl$_2$           &  279.3   & 80.0     & 2.7      &  18.5       & 279.3         &    80.0                 & 0.034    & 0.010 \\
    Graphene~\cite{PhysRevB.85.125428}   &  352.7   & 352.7    & 60.9     &  145.9      & 342.2         &    342.2                & 0.173    & 0.173
\end{tabular}
\end{ruledtabular}
\end{table*}

We further investigate the mechanical properties of monolayer MoOCl$_2$, including the elastic constants ($C_{ij}$), Young's Modulus ($Y$) and Poisson's ratio ($\nu$). Our results are listed in Table~\ref{tab: ec}.
One finds that the obtained elastic constants satisfy the mechanical stability criterion for rectangular lattices~\cite{PhysRevB.90.224104}:
\begin{equation}
  C_{11}C_{22}-C^2_{12} > 0,\qquad C_{66} > 0,
\end{equation}
which further demonstrates that the monolayer is stable. Again, strong anisotropy is observed in Young's modulus and Poisson's ratio for the two in-plane directions. Notably, Poisson's ratio of monolayer MoOCl$_2$ is much less than most 2D materials~\cite{PhysRevB.85.125428,PhysRevB.73.041402,ZHANG20114080}, implying that the material could be useful for making flexible devices.

\section{\label{ssec: bands} Electronic Structure}

\begin{figure}[tb]
    \includegraphics[width=0.48\textwidth,angle=0]{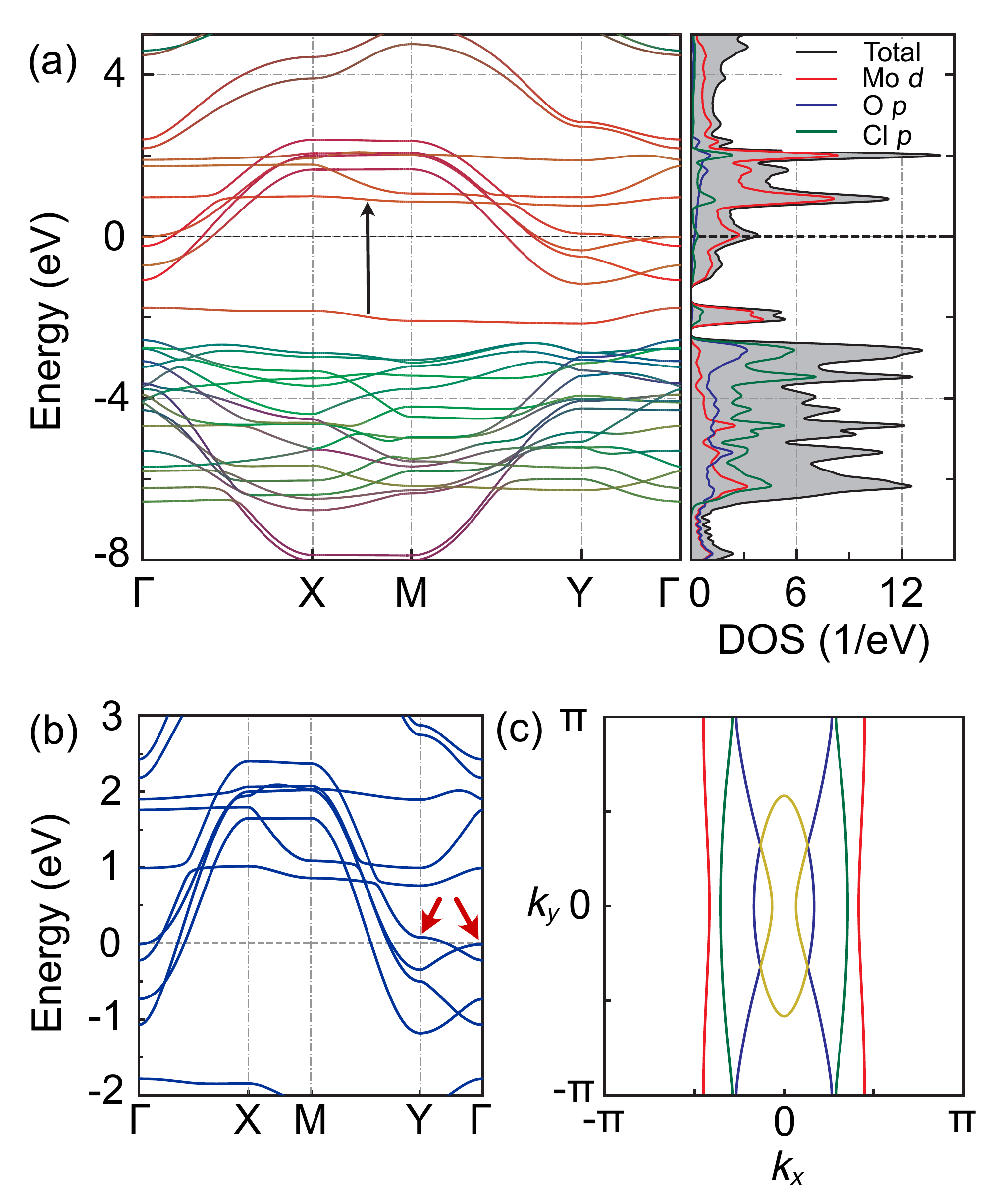}%
    \caption{\label{fig: bands} (a) Band structure and orbital projected DOS of monolayer MoOCl$_2$. The different orbital contributions are indicated by different colors. (b) shows the low-energy bands near the Fermi level. The arrows indicate two saddle points close to the Fermi level. (c) Fermi surface of monolayer MoOCl$_2$. }
\end{figure}

Next, we investigate the electronic band structure of monolayer MoOCl$_2$. The result obtained from our DFT calculation is plotted in Fig.~\ref{fig: bands}(a), along with orbital projected density of states (DOS). The effect of spin-orbit coupling (SOC) on the low-energy bands is found to be small, so the result shown here is without SOC (the result with SOC is presented in the Supplemental Material~\cite{Supplemental}). From Fig.~\ref{fig: bands}(a), one finds that monolayer MoOCl$_2$ is a metal. The low-energy states around the Fermi level are
mainly from the Mo-$4d$ orbitals. The O-$2p$ and Cl-$3p$ orbitals are located away from the Fermi level (below $-3.0$ eV).

Importantly, the band dispersion is highly anisotropic. There are four energy bands crossing the Fermi level. One can see that these bands are highly dispersive along the $x$ direction (e.g., $\Gamma$-$X$ and $M$-$Y$ paths). In contrast, they become quite flat along the $y$ direction (e.g., $X$-$M$ and $Y$-$\Gamma$ paths). This is consistent with our previous analysis that the monolayer is strongly coupled along the Mo-O chains, whereas the inter-chain coupling in the orthogonal direction is much weaker.

This strong anisotropy is also manifested in the Fermi surface geometry. Figure~\ref{fig: bands}(c) shows the calculated Fermi surfaces (contours). These contours are strongly elongated along the $k_y$ direction, as a result of the weak band dispersion along $y$. Notably, the strong anisotropy makes several contours cross the BZ boundary, giving rise to open orbits. The signature of such open orbits can be found in the magneto-transport, namely, the magneto-resistance can grow without limit when the applied magnetic field increases~\cite{ashcroft1976solid}. This non-saturation behavior has been observed in bulk MoOCl$_2$, which also has open orbits at the Fermi level along $k_y$~\cite{Wang.2020p6}.

Finally, we mention that there is a saddle point at $\Gamma$ very close to the Fermi level, as marked by the red arrow in Fig.~\ref{fig: bands}(b). Around the point, the band energy increases (decreases) along the $\Gamma$-$X$ ($\Gamma$-$Y$) direction. According to the $D_{2h}$ little group at $\Gamma$, the band dispersion around this point should be described by
\begin{equation}
    E(\boldsymbol{k}) = \varepsilon_0 + A_x k^2_x + A_y k^2_y,
\end{equation}
where $\varepsilon_0=-0.010$ eV is the energy of the saddle point. The two parameters $A_x$ and $A_y$ can be extracted from the DFT band dispersion. The results are $A_x=4.357$ eV$\cdot$\AA$^2$ and $A_y=-1.779$ eV$\cdot$\AA$^2$. Their opposite signs conform with the saddle-like band dispersion. This type of band dispersion is another reflection of the strong in-plane anisotropy. Another saddle point can be observed at $Y$ at energy 81 meV above the Fermi level. These saddle points contribute to the small peak in the DOS around the Fermi level.

\section{\label{ssec: trans} Hyperbolic optical property}

\begin{figure}[tb]
    \includegraphics[width=0.45\textwidth,angle=0]{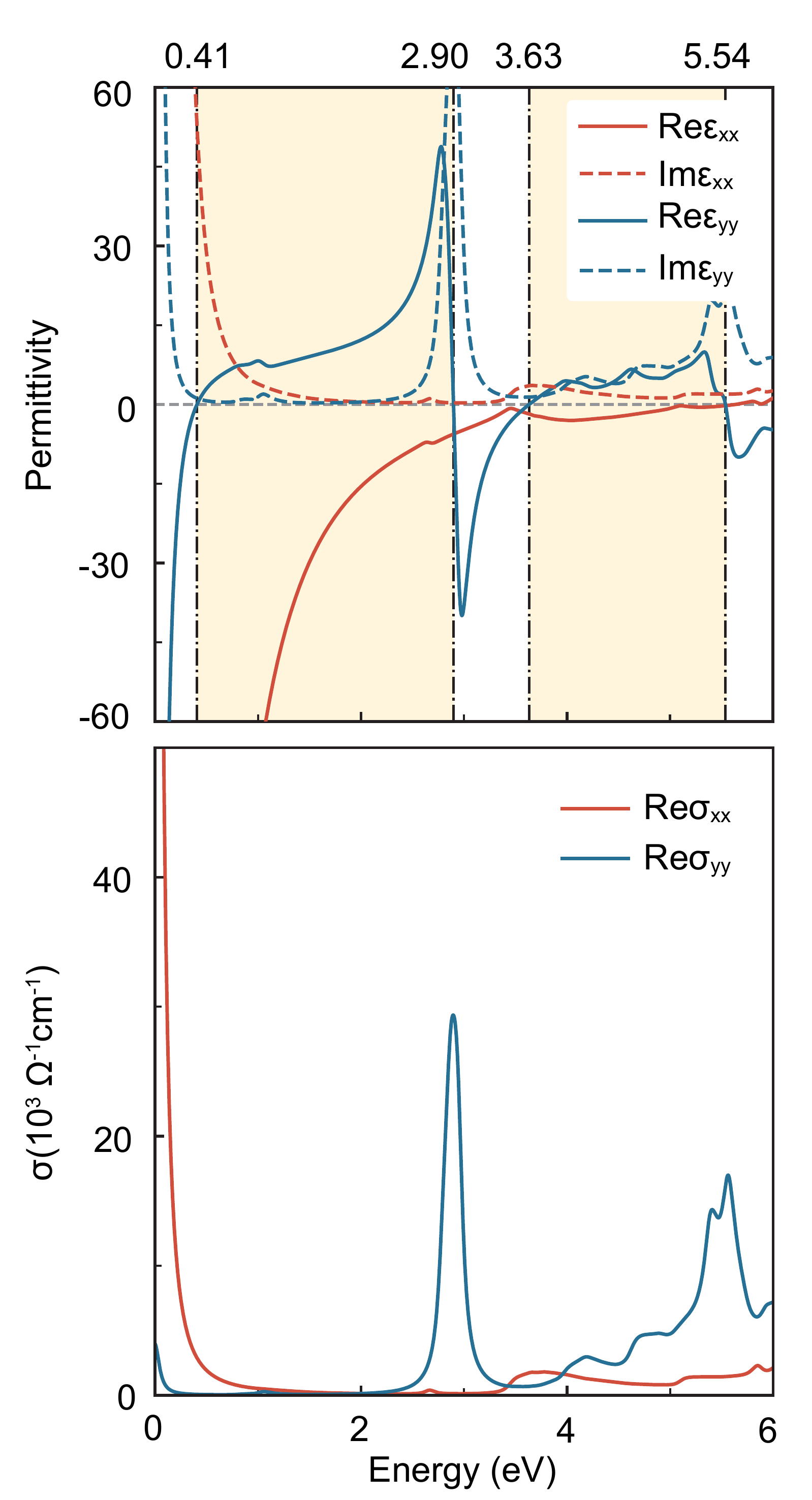}
    \caption{\label{fig: opt_cond} (a) Real and imaginary parts of permittivity $\varepsilon (\omega)$ of monolayer MoOCl$_2$. The shaded region indicates the hyperbolic frequency window. (b) Real part of the optical conductivity of monolayer MoOCl$_2$.}
\end{figure}

As we mentioned at the very beginning, anisotropic metals naturally possess hyperbolic permittivity. To understand this point, we note that from the Drude model, for simple metals, the real parts of the permittivity tensor elements take the form of
\begin{equation}\label{RE}
  \text{Re}\left[\varepsilon_{\alpha\beta}(\omega)\right]=1-\frac{(\omega^p_{\alpha\beta})^2}{\omega^2+\gamma^2},
\end{equation}
where $\gamma$ is the electron life-time broadening, and $\omega^p_{\alpha\beta}$ is the plasma frequency tensor element. The expression in Eq.~(\ref{RE}) is an increasing function of $\omega$. It takes negative value at low frequencies, and approaches $+1$ when $\omega\rightarrow\infty$. Typically, for good metals, $\omega^p_{\alpha\beta}\gg\gamma$, so $\text{Re}\left[\varepsilon_{\alpha\beta}(\omega)\right]$ crosses zero when $\omega\approx \omega^p_{\alpha\beta}$.

The plasma frequency tensor is an intrinsic band property,
given by
\begin{equation}\label{wp}
  (\omega^p_{\alpha\beta})^2=-\frac{4\pi e^2}{V}\sum_{n,\bm k} 2f'_{n\bm k}\left(\hat{\bm e}_\alpha\cdot\frac{\partial E_{n,\bm k}}{\partial\bm k}\right)\left(\hat{\bm e}_\beta\cdot\frac{\partial E_{n,\bm k}}{\partial\bm k}\right),
\end{equation}
where $f_{n\bm k}$ is the equilibrium distribution function, and $\hat{\bm e}_\alpha$ ($\hat{\bm e}_\beta$) is the unit vector along the $\alpha$ ($\beta$) direction. For our current 2D system with $D_{2h}$ point group symmetry, only the diagonal elements $\omega^p_{xx}$ and  $\omega^p_{yy}$ are non-vanishing. From (\ref{wp}), we see that each element is proportional to the squared band velocity along $x$ or $y$ averaged on the Fermi surface. Since we know from the band structure of monolayer MoOCl$_2$ that the dispersion along $y$ is much weaker than along $x$, we must have
\begin{equation}
\omega_{xx}^p\gg \omega_{yy}^p.
\end{equation}
It follows that $\text{Re}(\varepsilon_{xx})$ and $\text{Re}(\varepsilon_{yy})$ must cross zero at very different frequencies. Then, the frequency window between their zero points defines the hyperbolic window, in which $\text{Re}(\varepsilon_{xx})$ and $\text{Re}(\varepsilon_{yy})$ have opposite signs.

The above analysis is confirmed by our DFT calculation. In Fig.~\ref{fig: opt_cond}(a), we plot both real and imaginary parts of $\varepsilon_{\alpha\alpha}$ $(\alpha=x,y)$. In the calculation, we take $\gamma\sim 50$ meV, the value extracted from the transport measurement on the bulk sample at 10 K~\cite{Wang.2020p6}. From the result, one clearly observes that there are two wide hyperbolic windows from 0.41 eV to 2.90 eV and from 3.63 eV to 5.54 eV, in which \(\text{Re}(\varepsilon_{yy})\) is positive whereas $\text{Re}(\varepsilon_{xx})$ is negative. The former window is even wider than the previously reported one in Ca$_2$N (from 0.38 eV to 1.40 eV)~\cite{Guan.2017}, and it overlaps a large portion with the visible light spectrum. Moreover, one finds that the imaginary parts of the permittivity, which represent dissipation, are quite small for most part of this hyperbolic window (less than 1 from 1.58 eV to 2.30 eV). All these characters are much desired for applications. 2D materials with such ideal hyperbolic properties have not been reported before.

To take a closer look at optical absorption, in Fig.~\ref{fig: opt_cond}(b), we plot the real parts of the optical conductivities. They are connected with the imaginary parts of $\varepsilon_{\alpha\alpha}$, because of the fundamental relation
\begin{equation}
  \varepsilon_{\alpha\beta}(\omega)=1+\frac{4\pi i}{\omega}\sigma_{\alpha\beta}(\omega).
\end{equation}
One can see that because $\omega_{xx}^p$ is much larger than $\omega_{yy}^p$, $\text{Re}(\sigma_{xx})$ has a much stronger Drude peak (i.e., the peak close to zero frequency) than $\text{Re}(\sigma_{yy})$. However, $\text{Re}(\sigma_{xx})$ is strongly suppressed above 0.70 eV in Fig.~\ref{fig: opt_cond}(b). In contrast, $\text{Re}(\sigma_{yy})$ is suppressed between 0.09 eV and 2.54 eV, and it has a peak around 2.89 eV. By analyzing the band structure, we find this peak corresponds to the interband transitions between two almost parallel bands, as indicated by the arrow in Fig.~\ref{fig: bands}(a). This dramatic difference between  $\text{Re}(\sigma_{xx})$ and $\text{Re}(\sigma_{yy})$ indicates a strong linear optical dichroism, i.e., for incident lights with different linear polarizations, their absorption spectra will be very different.


\section{\label{ssec: strain} Strain effect}

\begin{figure}[!ht]
    \includegraphics[width=0.45\textwidth,angle=0]{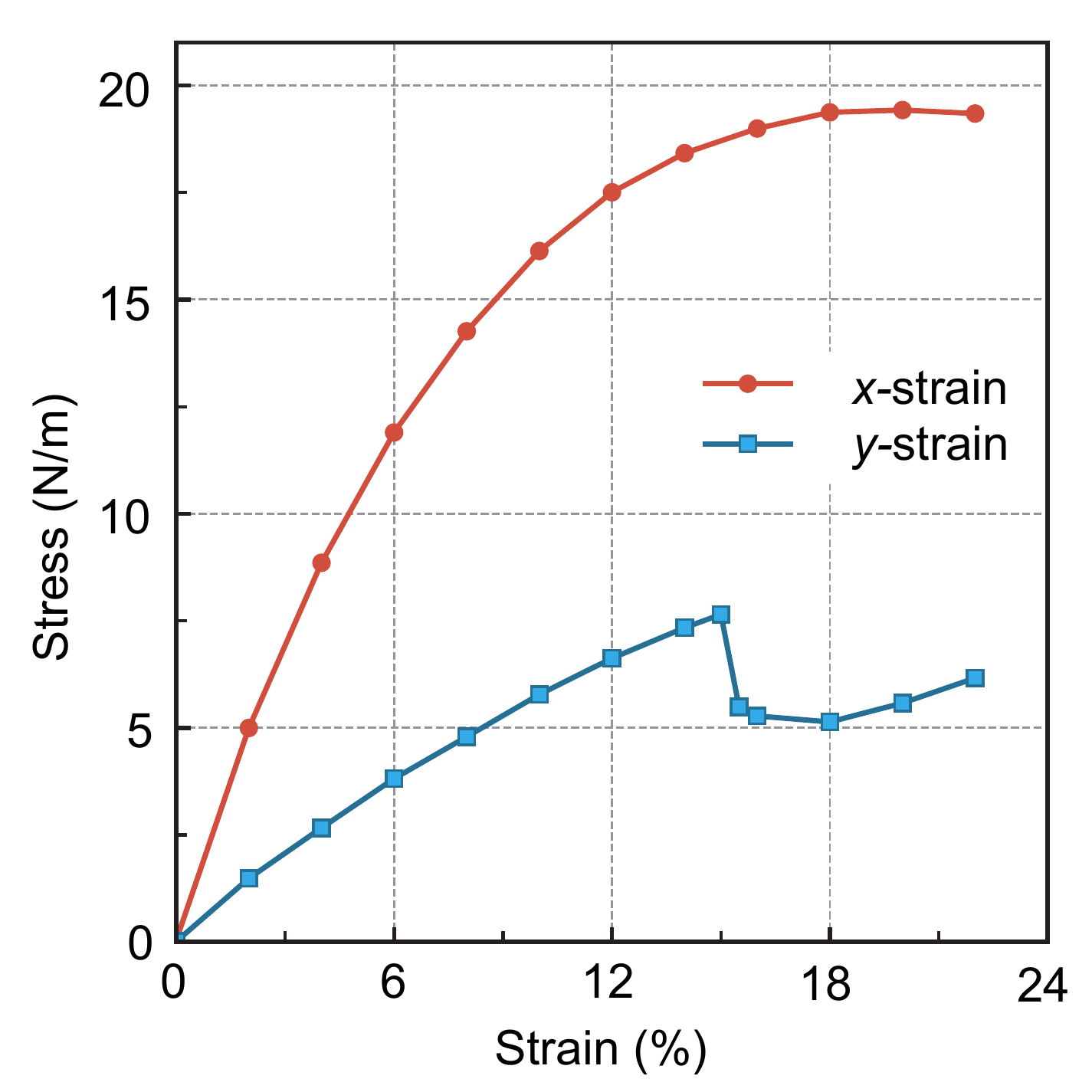}%
    \caption{\label{fig: strain} Strain-stress curves for monolayer MoOCl$_2$ for strains applied in $x$ and $y$ directions. }
\end{figure}

2D materials can usually sustain very large lattice strains ($>10\%$). And strain has been proved to be an effective method to tune the physical properties of 2D materials. Here we calculate the strain-stress relations for monolayer MoOCl$_2$, and the result is plotted in Fig.~\ref{fig: strain}. One observes that the curves for strains along $x$ and $y$ directions are quite different, again reflecting the large anisotropy. The critical strain for the $x$ direction is about 18\%, and for the $y$ direction is about 15\%. The large values of the critical strain demonstrate that monolayer MoOCl$_2$ is quite flexible.

\begin{figure}[!h]
    \includegraphics[width=0.47\textwidth,angle=0]{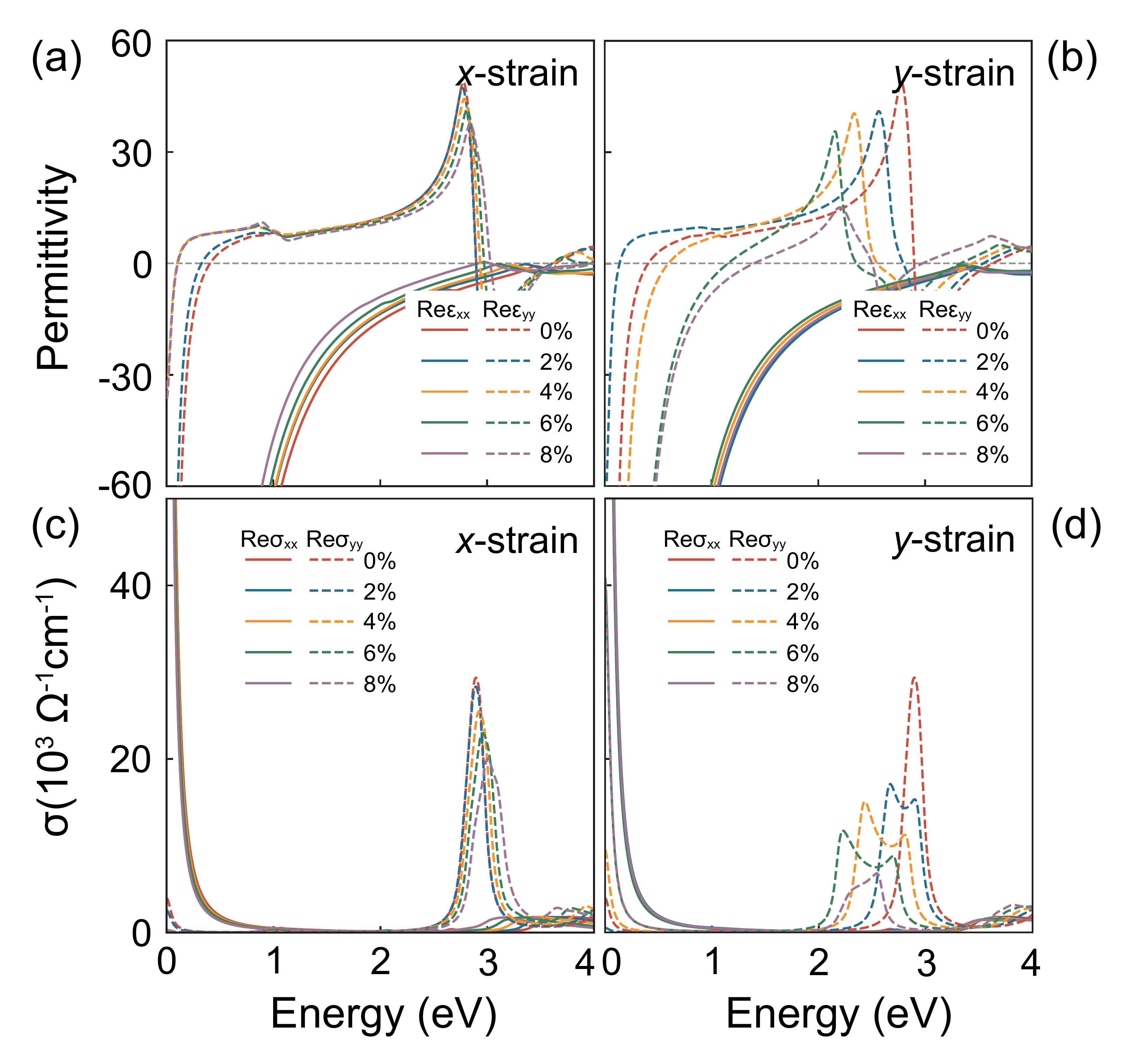}%
    \caption{\label{fig: opt_strain} Effects of strain on (a,b) $\text{Re}\left[\varepsilon (\omega)\right]$ and (c,d) optical conductivity $\text{Re}\left[\sigma (\omega)\right]$. (a) and (c) are for strains along $x$. (b) and (d) are for strains along $y$.}
\end{figure}

Then we consider the effect of strain on the physical properties of monolayer MoOCl$_2$. We shall focus on the hyperbolic optical response below 4.0 eV, as it is the most interesting feature here. (The change in the band structure and the Fermi surface geometry is presented in the Supplemental Material~\cite{Supplemental}.) Figure~\ref{fig: opt_strain}(a-b) shows the calculated $\text{Re}(\varepsilon_{\alpha\alpha})$ versus the applied uniaxial strains. Here, to simulate the stretching of a 2D sample, the strain is applied along $x$ or $y$ direction, and the structure is fully relaxed in the orthogonal direction. One observes that for tensile strains applied along $x$ (Fig.~\ref{fig: opt_strain}(a)), the hyperbolic window is gradually shifted to lower frequencies. Comparing the unstrained case and the 8\% strained case, the hyperbolic window is red shifted from $(0.41, 2.90)$ eV to $(0.10, 3.03)$ eV. Meanwhile, for strains applied along $y$ (Fig.~\ref{fig: opt_strain}(b)), the change is more pronounced. While the change in $\text{Re}(\varepsilon_{xx})$ is relatively small, the change in $\text{Re}(\varepsilon_{yy})$ is quite large. The $\text{Re}(\varepsilon_{yy})$ curve and its zero point first undergo a red shift at small strains, then they revert to a blue shift for strains greater than 4\%. The hyperbolic window changes from $(0.41, 2.90)$ eV at zero strain to $(1.40, 2.50)$ eV at 8\% strain. Thus, there is a range of frequencies, e.g., between 0.41 eV and 1.40 eV, in which the system is hyperbolic at zero strain but changed to elliptic at 8\% strain. It follows that by tuning the applied strain, we can completely change the optical character of the material. This is a great advantage for applications.

In Fig.~\ref{fig: opt_strain}(c,d), we also plot the variation of $\text{Re}(\sigma_{\alpha\alpha})$ versus the applied strain. The anisotropy (linear dichroism) remains large for these applied strains. Similar to the trend in Fig.~\ref{fig: opt_strain}(c), the change is more pronounced for strains applied along the $y$ direction. In Fig.~\ref{fig: opt_strain}(d), one observes that the single interband peak at 2.89 eV starts to split into two and red shift with increasing strain. These results will be useful for tuning the optical absorption of the material and for probing the strain via optical detection.

\section{\label{sec: conlusion} Discussion and Conclusion}

\begin{figure}[!h]
    \includegraphics[width=0.45\textwidth,angle=0]{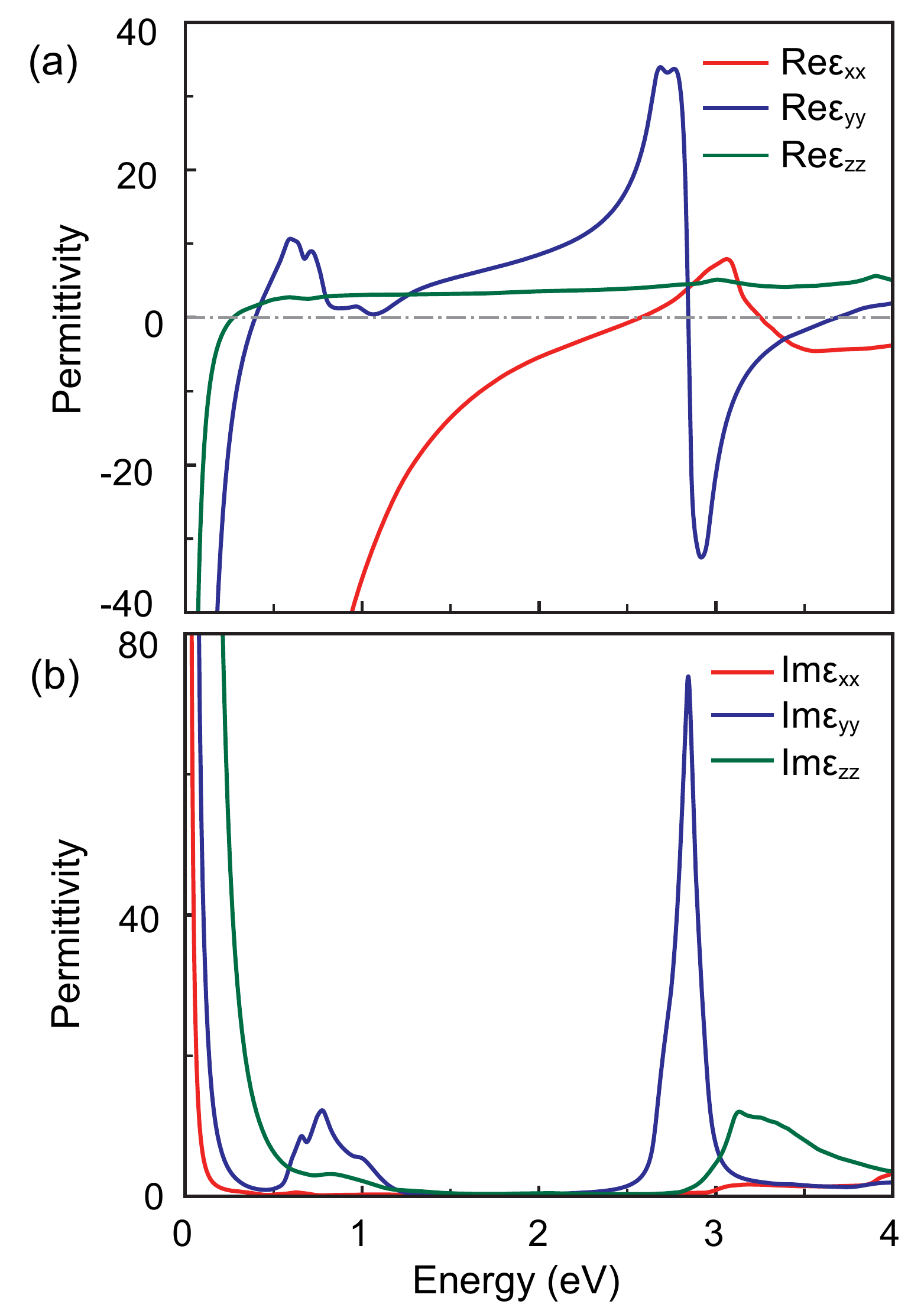}%
    \caption{\label{fig: epsilon_bulk} (a) Real and (b) imaginary parts of permittivity tensor elements for bulk MoOCl$_2$.}
\end{figure}

We have revealed monolayer MoOCl$_2$ as a highly anisotropic 2D metal. Such a character is quite unique among the existing 2D materials. Previously, the anisotropic feature has been mainly discussed in black phosphorene and related materials. However, these materials are intrinsically semiconductors, and their degree of anisotropy is not as high as monolayer MoOCl$_2$. From our discussion, it is clear that the metallic character is important for its excellent hyperbolic optical property.

Besides the properties discussed here, the anisotropy will manifest in many other properties, such as the plasmonic waves, charge and thermal transport, and etc. And for 2D materials, there are many well developed approaches to tune their properties, such as the chemical modification, doping by adsorption, substitution, or gating. They may have pronounced effect on monolayer MoOCl$_2$ and change its optical response.
We hope our current work can motivate more subsequent studies on this interesting 2D material.

Finally, regarding the hyperbolic optical property, we note that bulk MoOCl$_2$ is also an excellent candidate among 3D materials.  The band structure of bulk MoOCl$_2$  has been studied in Ref.~\cite{Wang.2020p6}. Here, we calculate its permittivity and the result is shown in Fig.~\ref{fig: epsilon_bulk}. One can see that there is also a large hyperbolic window between 0.39 eV and 2.57 eV, in which $\text{Re}(\varepsilon_{xx})$  is negative, whereas the other two are positive. And Fig.~\ref{fig: epsilon_bulk}(b) shows that the dissipation in the range 1.17 eV to 2.43 eV of this hyperbolic window is very small. Thus, bulk MoOCl$_2$ could serve as a high-performance 3D hyperbolic material.

In conclusion, we have proposed a new 2D material---the monolayer MoOCl$_2$. We predict that the monolayer could be easily obtained from the existing bulk sample by mechanical exfoliation. We have systematically investigated its physical properties, and revealed the material as a highly anisotropic 2D metal. The anisotropy originates from the underlying lattice structure, and manifests in a wide range of properties. The most prominent one uncovered here is the hyperbolic optical property. We show that the material hosts two wide hyperbolic windows from 0.41 eV and 2.90 eV and from 3.63 eV to 5.54 eV. The former window overlaps a large part of the visible spectrum and has very little dissipation. Such a characteristic is superior than all existing 2D materials, to our best knowledge. In addition, we demonstrate that the optical property can be effectively tuned by strain, and even a transition between hyperbolic and elliptic characters can be achieved. Our findings establish monolayer MoOCl$_2$ as a unique addition to the 2D materials family. Its fascinating properties will complement other 2D materials and hold great promise for nanoscale electronic, mechanic, and optical applications.


\begin{acknowledgments}
    The authors thank Si Li, Shan Guan, Z. L. Liu, and  D. L. Deng  for valuable discussions. This work was supported by the National Natural Science Foundation of China (No.~11604273), and the Singapore Ministry of Education AcRF Tier 2 (MOE2019-T2-1-001).
\end{acknowledgments}

\bibliographystyle{apsrev4-1}
\bibliography{moocl2}

\end{document}